# Device Considerations for Nanophotonic CMOS Global Interconnects

Sasikanth Manipatruni, Michal Lipson, *Senior Member, IEEE* and Ian A. Young, *Fellow, IEEE*

*Abstract*— We introduce an analytical framework to understand the path for scaling nanophotonic interconnects to meet the energy and footprint requirements of CMOS global interconnects. We derive the device requirements for sub 100 fJ/cm/bit interconnects including tuning power, serialization-deserialization energy, and optical insertion losses. Using CMOS with integrated nanophotonics as an example platform, we derive the energy/bit, linear and areal bandwidth density of optical interconnects. We also derive the targets for device performance which indicate the need for continued improvements in insertion losses (<8dB), laser efficiency, operational speeds (>40 Gb/s), tuning power (<100 µW/nm), serialization-deserialization (< 10 fJ/bit/Operation) and necessity for spectrally selective devices with wavelength multiplexing (> 6 channels).

## I. Introduction: A framework for scaling CMOS nanophotonic global interconnects

Increasing computational demands of enterprise and datacom (DC) applications [1, 2] have created a need for scalable interconnect solutions for high performance computing (HPC). While the present industry focus is on the adoption of inter-chip optical interconnections [3, 4]; the rapid adoption of multicore processors in DC and HPC [5] with high demands on bandwidth density and efficiency [1] may necessitate new interconnect solutions for same-die global interconnects [6-9]. Given the rapid progress in CMOS compatible nano-photonics using III-V [10], Germanium [11] as well as Silicon based [10-16] platforms, the on-chip adaptability of optical interconnects for global wires [17] needs to be revisited.

In this paper, we develop a systematic framework for scaling nanophotonic interconnects by using device and system level arguments. We use CMOS with integrated nanophotonic devices as an example platform but the analytical framework can be applied to other platforms [e.g. 10, 11]. The device advances in couplers [18], low loss waveguides [19], modulators [20-24], switches [25-28], multi-wavelength devices [29-30] & detectors [31-35] can be put in context with the targets for on- chip integration using this framework.

We derive the total interconnect energy per bit, areal bandwidth density and linear bandwidth density for a silicon photonic link considering the device parameters. We arrive at a minimal set of features for nanophotonic devices for building a scalable on chip photonic network. We note that we limit our analysis to how photonic devices can be scaled to meet on-chip interconnect energy/bit and bandwidth density requirements. We compare the energy/bit/mm, linear bandwidth density of the optical interconnect with generic interconnect targets for CMOS. A direct comparison with a future advanced low swing voltage (LSI) electrical on-chip interconnects is hard to achieve within the scope of the paper since such an analysis has to fundamentally comprehend the variability limits to LSI interconnects [59, 60].

## II. Figures of merit for nanophotonic interconnects

We discuss four critical figures of merit for nanophotonic interconnects based on physical constraints of the optical and electrical properties of a silicon based material system. Namely, a) Energy consumption per bit (E) b) Interconnect density (β) c) Single channel bandwidth (f) d) Areal bandwidth density (D).

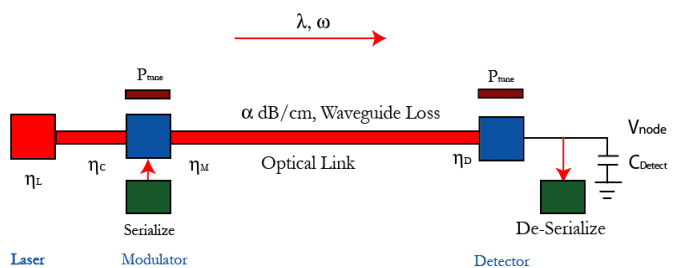

Figure 1: A minimal nanophotonic link with an optical source, couplers, modulators, waveguide and a detector. A serializer and deserializer are considered to obtain the optimum operating speed of the link. Tuning at both ends is assumed to operate the link at a specific wavelength.

## III. Energy/bit of a nanophotonic interconnect (E)

We will derive the minimum bound for an optical interconnect electrical energy per bit considering the performance of the modulators, detectors, waveguide and coupling insertion losses. For the following analysis, we have assumed a receiver less topology for optical interconnect as proposed in Miller et al [36]. While, this is not the optimal optical link design for all operating conditions (see Appendix A, B), we believe this provides reasonable direction for the optical device requirements when the on-chip detector capacitance is low [36, 37]. The total optical interconnect energy per bit can be

Manuscript received XXXXXXXXXXX. S. Manipatruni, and I. A. Young are with Components Research, Intel Corp., Hillsboro, OR 97124, USA (e-mail: sasikanth.manipatruni@intel.com) M. Lipson is at School of Electrical and computer engineering, Cornell University, Ithaca, NY 14850 & Kavli Institute at Cornell, Ithaca, NY 14853, USA

written (in the absence of tuning power and serialization) as a sum of energy from the source and the electro-optic modulator's energy as:

$$E_{total} = E_{Source,Detect} + E_{EO} \quad (1)$$

Where $E_{Source\text{-}detect}$ is the energy spent in the source laser and the detector energy; $E_{EO}$ is the energy spent in electro-optic coding of the electrical information into an optical signal. A lower bound to the interconnect energy can be obtained by assuming that the detector needs to charge a capacitor of capacitance $C_d$ to a voltage $V_r$ corresponding to a specific CMOS node [36]. While this is an aggressive requirement, this assumption lets us derive a minimum bound for energy per bit requirements. $E_{source,Detect}$ can be written in terms of drive laser parameters and insertion losses as

$$E_{Source,Detect} > \frac{\hbar\omega}{\eta_L \eta_D \eta_M \eta_C} \cdot \frac{V_r C_d}{e} \cdot 10^{\frac{\alpha L}{10}} \quad (2)$$

where $V_r$ is the minimum voltage to which the detector capacitance is to be charged, $\eta_L$, $\eta_D$ are the quantum efficiencies of the laser and detector normalized to the maximum values, $\eta_C$ is the laser to waveguide coupling efficiency, $\eta_D$ includes the waveguide to detector coupling efficiency. $\eta_M$ is the modulator insertion loss, $\alpha$ is the insertion loss of the waveguides in dB/cm, L is the length of the interconnect in cm. The above equation is a reasonable approximation for the following conditions: a) the detector RC response is significantly faster than the optical pulse width b) the received optical power & extinction ratio exceeds the bit error rate requirement (see appendix B) of the link and c) the collected optical power at the receiver is always adjusted to allow full voltage at the detector. We also note that an on chip receiver drives a significantly lower load capacitance (a few transistor gate capacitances on the order of aFs).

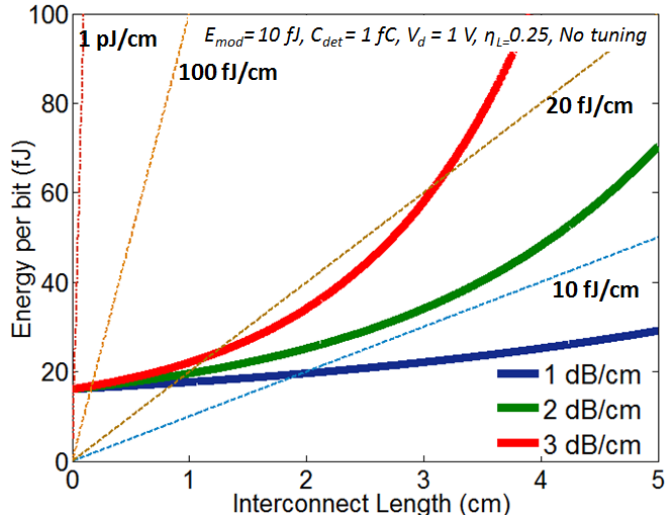

Figure 2 : Idealized interconnect energy/bit assuming no thermal tuning and compact modulators, detectors; 1 dB coupling loss, 1 dB modulator insertion loss, -1 dB detector efficiency are assumed. Dotted lines show fixed energy/bit/length points.

The minimum electro-optic conversion energy per bit ($E_{EO}$) is arrived at using the modal volume of the modulator and the injected charge density for a given transmission change. We assumed a modulator drive voltage $V_m$, electro-optic modal volume $\Theta$, the optical transmission change $\Delta T$. $dT/dn$ is the spectral sensitivity of the optical device. $dn/d\rho$ is change in refractive index (n) vs. carrier concentration ($\rho$) in the electro-optic device.

$$E_{EO} > \frac{V_m \Theta \Delta T}{\left(\frac{dn}{d\rho}\right)\left(\frac{dT}{dn}\right)} \quad (3)$$

We show that an idealized nanophotonic interconnect in the absence of tuning power & electrical I/O overheads can achieve sub 100 fJ/bit/cm operation. Modulator switching energy approaching 10 fJ/bit can be expected in the near future in the depletion based & ultra-low modal volume modulators [23, 38]. Figure 2, shows the energy vs. distance scaling of a nanophotonic interconnect with $E_{mod}$=10 fJ/bit modulation energy, $C_d$=1 fF detector capacitance, 1 dB coupling loss, 1 dB modulator insertion loss, -1 dB detector efficiency and 25 % efficiency laser source. (See Appendix C)

*A. Effect of laser efficiency on the energy per bit*

The power efficiency of the laser has a significant effect on the interconnect energy per bit. In figure 3 we show the interconnect energy per bit for varying laser efficiency (defined as optical output power vs. electrical power supplied to the laser). The low inefficiency of the laser may arise due to several factors including the requirement for thermoelectric cooling, collection efficiency & leakage power. At 5 % wall plug efficiency the interconnect energy/bit at 1 cm length can approach 50 fJ/bit/cm, for idealized interconnects with no tuning requirement. The effect of additional insertion loss due to routing and selective devices is described in Appendix E.

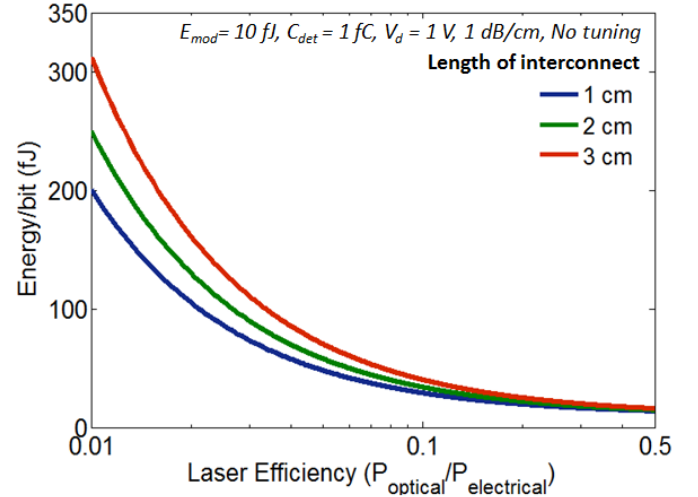

Figure 3: Effect of laser efficiency on the interconnect energy/bit in an idealized interconnect with no thermal tuning;

*B. Effect of tuning nanophotonic devices to offset variability & temperature dependence*

We show that higher operating speeds of the devices may allow for the averaging of the tuning power required over many bits in order to achieve low energy per bit. Tuning of nanophotonic devices is essential due to the intrinsic temperature dependence of refractive index of solid state materials, wafer level variability, with run time operating temperature variability [39]. The total power including the tuning power for modulator and detector wavelength selective devices can be written as

$$E_{total} > \frac{\hbar\omega}{\eta_L \eta_D \eta_M \eta_C} \cdot \frac{V_r C_d}{e} \cdot 10^{\frac{\alpha L}{10}} + \frac{V_m \Theta \Delta T}{\left(\frac{dn}{d\rho}\right)\left(\frac{dT}{dn}\right)} + \frac{2}{B}P_{tune} \times \Delta\lambda$$

(4)

Where we included the tuning power per nanometer of correction $P_{tune}$ to correct the operating wavelength of the modulator & detector by $\Delta\lambda$. B is the bit rate of the link. In figure 4, we show the effect of the tuning power on the total interconnect energy. The constant power penalty due to tuning will mandate operation at higher speeds so that the tuning power can be shared among more bits per second.

Higher operating speeds of interconnects will be necessary to achieve an energy/bit below 100 fJ/bit/cm since the tuning power imposes a significant constraint on the energy efficiency of nanophotonic interconnects. As shown in figure 4, 100 fJ/bit energy targets can be reached only at 40 Gb/s when a 2 nm (20 C) correction is required. The run time temperature control for the micro-processors is expected to be 20 C with a spatial variation of 50 C in temperature [39]. Hence significant advances, in temperature independent device operation [40] or highly efficient low overhead tuning schemes remain to be developed [41, 42]. We note that packaging and module level cooling may significantly change the tuning requirements.

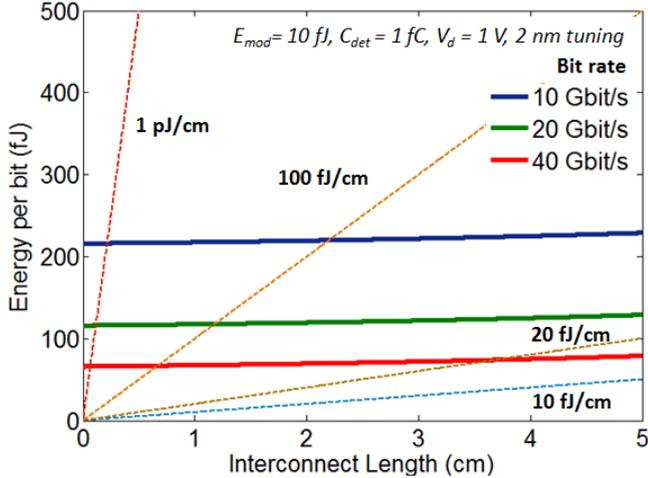

Figure 4: Effect of tuning power on the interconnect energy/bit; assuming a 100 μW/nm tuning [e.g. 42] mechanism for transmitter and detector with 20 K tuning requirement.

*C. Effect of on-chip serialize-deserialize operations*

We show that efficient electrical serialize and deserialize operations are essential to operate the optical links at higher operating speeds. We obtain the optimum operation speeds of the silicon optical interconnect by including the energy cost of serialize-deserialize operations and the tuning power.

We modeled the power penalty for serialize and deserialize (SerDes) operations as a constant energy per bit per serialization order. The total energy of the link can be written as:

(5)

$$E_{total} > \frac{\hbar\omega}{\eta_L \eta_D \eta_M \eta_C} \cdot \frac{V_r C_d}{e} \cdot 10^{\frac{\alpha L}{10}} + \frac{V_m \Theta \Delta T}{\left(\frac{dn}{d\rho}\right)\left(\frac{dT}{dn}\right)} + \frac{2}{B}P_{tune} \times \Delta\lambda + E_{SD}\frac{B}{2F_{clock}}$$

where $F_{clock}$ is the system clock, $E_{SD}$ is the energy per bit per serialization order (N). The SerDes are used for scaling the bit rates beyond twice the system clock. The exact functional form for the SerDes operations can be different, however, it is commonly understood that the higher the bit rate and degree of serialization, the larger is the energy for serializing and de-serializing. In figure 5, we show the effect of serialize, de-serialize power on the total energy per bit. Some recent examples of optimization for on-chip serial link SerDes are [52, 53]. For a large SerDes energy of 50 fJ/bit per serialization order, we see that the minimum of the energy is obtained when no serialization takes place at 2* $F_{clock}$ bit rate. However, for a lower SerDes energy (10 fJ/bit), the penalty due to SerDes is not significant enough to change the behavior of the interconnect energy. The minimum energy is then obtained when the interconnect is operated at the maximum possible drive conditions. (See Appendix D for SerDes energy scaling with CMOS technology node).

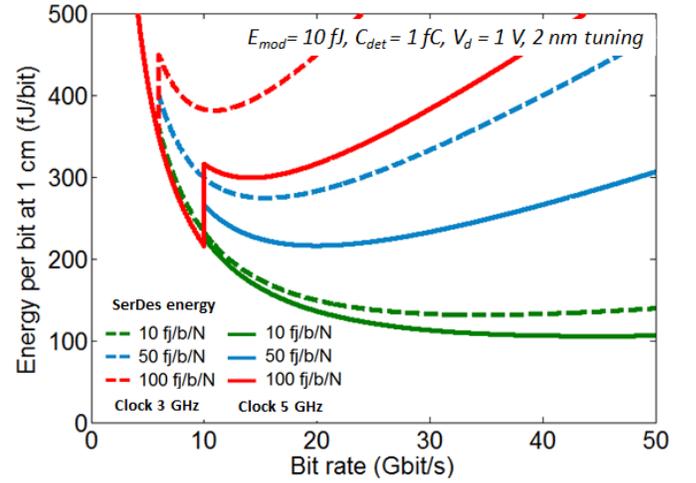

Figure 5: Effect of Serialize & Deserialize (SerDes) operations on the interconnect energy/bit; SerDes is employed for Bit rate > 2x Fclock;

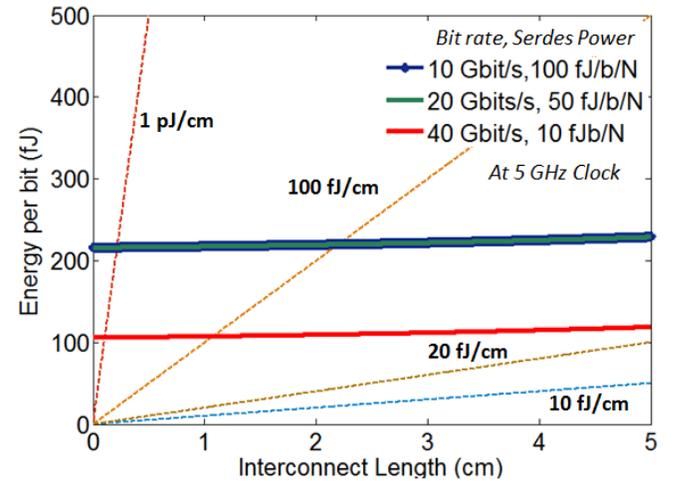

Figure 6: Total Energy of the optical interconnects vs. length. Intercept points with various energy/bit/length are shown. Bit rates and SerDes energy corresponding to the minima in figure 6 are used for the example cases.

We also study the effect of the system clock on the behavior of the total interconnect energy per bit considering tuning power, serialization as well as device insertion losses.

The energy penalty due to serialization can be minimized by operating at the highest available system clock. We also assumed that a distributed clock is available throughout the chip. The clock distribution from the local source to the SerDes is considered local distribution and is ignored. We see that at a 5 GHz system clock, with a SerDes power of 10 fJ/bit/Operation and a tuning power of 100 µW/nm, 150 fJ/bit operation can be achieved for all bit rates above 20 Gb/s.

*E. Total Interconnect Energy Dependence on Length:*

We study the total optical interconnect energy as a function of length including insertion losses, laser, modulator and detector efficiency in figure 6. Cross over points of the optical interconnect energy/bit vs. generic interconnects with a fixed energy/unit area are shown in figure 6. A high energy/bit interconnect such as a 1pJ/cm interconnect [43] (for e.g. a full swing interconnect with a swing voltage of 0.68 V (ITRS 2011_ORTC-6, $V_{dd}$ for high performance) & Capacitance of 140 aF/µm (ITRS Table 2011_INTC2, 2020) will have cross over points as low as a few mm. However, an energy efficient interconnect with 100 fJ/cm [44, 59] will have a longer cross over point. It remains to be seen if the emerging electrical interconnects can meet the on chip bit error rate & variability requirements [59, 60] given the high aggregated bandwidth of microprocessors [61]. We believe that given the number of interconnects and the aggregated bandwidth in the microprocessor application of interconnects, error correction will be limited due to latency area and power considerations.

## IV. LINEAR INTERCONNECT BANDWIDTH DENSITY OF A NANOPHOTONIC INTERCONNECT ($\beta$)

Linear Bandwidth Density (LBD) of an interconnect is the bandwidth (B in bits/µm.s) of an interconnect normalized for the width of the interconnect. The interconnect density on a microprocessor scales as the wire pitch of interconnects scale as per ITRS requirements.

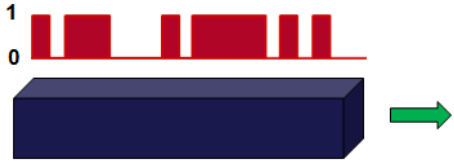

Figure 7: Bandwidth density: The bandwidth density of the waveguide is given by the aggregate data rate in the waveguide divided by the separation of two consecutive waveguides in an array of waveguides.

The fundamental limit to optical interconnect density is greatly enhanced by the high central carrier frequency and the ability to multiplex a large number of wavelengths [45]. For a nanophotonic waveguide array comprised of waveguides of width W, separated in a pitch of P, the bandwidth density (per micron) can be written as:

$$\beta_{Optical} = \frac{NB}{P} = NB\left(0.12\log_e\left(\frac{56.6L}{\pi}\right)\right)^{-1} \quad (7)$$

Where N is the number of WDM channels, B is the single channel bandwidth, P is the waveguide pitch and L (in microns) is the cross talk distance in microns. The pitch is the waveguide center to center pitch calculated for 250 nm (height) X 450 nm (width) waveguides such that a 3 dB coupling to the closest waveguide takes place for TE mode over a length of L (in microns) [46]. Novel CAD methods and wavelength allocation methods to separate the waveguides can reduce the effective pitch. Note that unlike the electrical case, the optical bandwidth density is not a strong function of the length of propagation. The dispersion effects enter the analysis as a secondary effect over several meters of propagation [47] enabling 1 Tb/s on a waveguide using WDM [45], thus indicating bandwidth density limits exceeding $10^{12}$ bits/µm.s.

*A. Length dependence of interconnect linear bandwidth density*

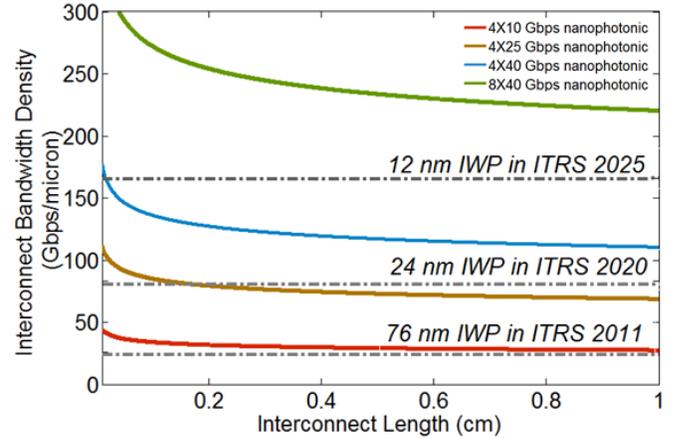

Figure 8: Interconnect density of optical WDM links and global wires. With ITRS targets for intermediate wires. IWP: intermediate wire pitch

Typical ITRS projections for electrical interconnect density at intermediate lengths are in the order of 20-200 Gb/s.µm. Given the scaling trends for the intermediate wires from 76 nm (2011) to 24 nm (2020) the electrical wires will increasingly be limited in BW density for longer distances (100 µm to 500 µm, arising from electromagnetic interference etc). Figure 8 shows the LBD for optical WDM waveguides plotted with a benchmark 1 µm intermediate wire interconnect at 100 Gb/s/µm. For >150 Gb/s. µm over global/intermediate distances (up to cm) a 8X40 Gb/s WDM will be essential

*B. Considerations on scaling the number of channels using micro-resonators*

Here we analyze two critical design considerations for scaling the bandwidth density using WDM: a) the channel spacing b) the total number of channels set by cavity free spectral range. We use 1st order optical micro-ring resonators as example resonators. We note that in general variety of micro-resonators and higher order designs can be employed. The wavelength spacing between the resonators can be controlled by considering the effect of waveguide and material dispersion. The functional dependence of resonance position of the rings can be given by:

$$\lambda_k = \frac{2\pi(r + \delta r(k))n_{eff}(\lambda_k)}{rn_{eff}(\lambda_0)}\lambda_0 \quad (8)$$

Where $\lambda_k$ is the position of the optical resonance of the k[th] micro-ring, r is the radius of the base micro-ring resonant at $\lambda_0$ is the radius perturbation introduced in the k[th] ring. We note that for a WDM microring bank spanning several 10s of nm

δr(k) will be a non-linear spacing variation obtained by including the variation in $n_{eff}$ ($\lambda_0$+ δλ. k) due to strong waveguide dispersion of high index contrast systems [45], waveguide bending and the material dispersion of the media. The channel spacing is also affected by the amplitude and phase cross talk due to off resonant interaction with the adjacent channels.

A second consideration is the free spectral range of the resonators to enable a large wavelength range for packing the WDM channels. The maximum number of channels that can be packed in a WDM system using micro-rings of radii $r + \delta r(k)$ with uniformly spaced channels at spacing δλ is given by

$$N = \frac{\Delta}{\delta\lambda} = \frac{\lambda_0}{\delta\lambda}\left(1 - \frac{m}{1+m}\frac{n_{eff}(\lambda_0 + k\delta\lambda)}{n_{eff}(\lambda_0)}\right) \quad (9)$$

where floor (N) is the number of channels, Δ is the free spectral range in wavelength, $m = 2\pi r n_{eff}(\lambda_0)/\lambda_0$ is the mode order for the base micro-ring. For example a micro-ring resonator of 1.5 micron radius can have an FSR of 62 nm allowing a large number of WDM channels [62]. One can see that a considerable design space is available using micro-resonators to meet the linear bandwidth density requirement.

### V. SINGLE CHANNEL BANDWIDTH OF A NANOPHOTONIC INTERCONNECT (F)

The limit to single channel bandwidth is decided by the operation speed of the receiver and transmitter. The fundamental limits to the electro-optic device speed are given by free carrier response times [20-23, 36] or electro-optic material response time or the driving capacitor time constant [15]. For photo-detectors and free carrier dispersion modulators:

$$f_{EO/OE} < \frac{1}{\tau_{min}} = \frac{v_{sat}}{nw} \quad (10)$$

Where $v_{sat}$ is the saturation velocity of carriers in silicon (set by the optical phonon dispersion), typical values of ~$10^7$ cm/s (for Si, Ge and III-Vs), w=λ/15~ 103 nm is space rate of decay of the evanescent field of the waveguide [50] and n is the arbitrary factor chosen such that $e^{-n}$ gives the factor by which the evanescent field decays. The typical clearance for placing thin film planar doped regions next to nanophotonic waveguides can be estimated to be 3λ/15 ~ 310 nm.

The switching speed of a scaled electro-optic device driven by a scaled single stage digital logic driver is [54]:

$$f_{Drive} < f_{EO/OE} = \left(\frac{C_n V_n}{I_n}\left(3\frac{I_{mod}}{I_n}+1.5\right)\right)^{-1} \quad (11)$$

where $C_n$, $V_n$, $I_n$ are the capacitance, voltage and current of a minimum sized transistor at a given technology node, $I_{modulator}$ is the peak current through the modulator. We plot the maximum switching speed of the direct logic drive as a function of the drive current for the modulator in Fig. 9. Gate lengths, voltages and delays are taken from ITRS HPC PIDS [1]. The voltage and current drive requirements for the EO devices therefore should be compatible with scaled CMOS for high speed operation. The voltage and current drive requirements for the EO devices therefore should be compatible with scaled CMOS for high speed operation.

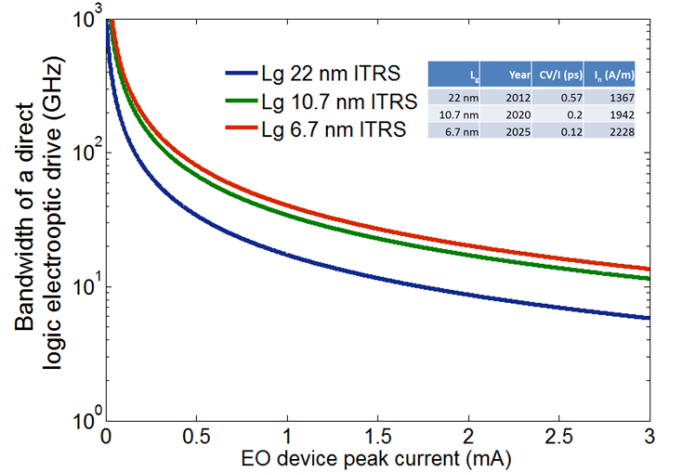

Figure 9: Bandwidth of a direct logic driven electro-optic nanophotonic device for scaled CMOS nodes.

### VI. AREAL BANDWIDTH DENSITY OF NANOPHOTONIC COMPONENTS (D IN BITS/MM$^2$)

Areal bandwidth density (ABD) of nanophotonic (transmitters/receivers) is the bandwidth generation/receiving capacity of components divided by the area of the device. The area taken by the wires and waveguides themselves is separately accounted for in the prior, interconnect bandwidth density metric. The role of ABD is to quantify the footprint taken by optical components to provide a certain bandwidth capacity. The modal volume of modulators as well as detectors enhanced by resonance effects are ultimately limited by diffraction limits

$$D_{optical} \equiv \frac{f}{Area} < \frac{v_{sat}}{nw}\left(\frac{\lambda}{2N}\right)^{-2} \quad (12)$$

N is the index refractive index of the guiding medium. The density will have to be adjusted to allow for the driver and receiver circuits (as shown by the driver scaling in section V). A 1.5 μm radius modulator operating at 10 Gb/s will reach bandwidth density of 1400 Tbit/s.mm$^2$ [38]. Improved speed, 3D integration and ultra-small modal volumes may be necessary for meeting the CMOS areal bandwidth density requirements.

### VII. DEVICE REQUIREMENTS FOR SCALABLE NANOPHOTONIC INTERCONNECTS

Based on the figures of merit proposed earlier, we present a minimal set of optical device requirements for replacing CMOS global interconnects. However, we note that specific device requirements derived above are for a single direct link and not a networked topology [6-9]. Four minimal features to enable optical components on chip are:

**A. High bandwidth, Broadband devices**: Higher speed of operation will allow large interconnect densities and offset the tuning power to reduce the energy/bit. Target speeds are in 10 to 40 Gbps for modulators with switch bandwidths to allow switching of 40 Gb/s signals.

**Table 1: Figures of merit for nanophotonic interconnects**

| FOM | Nanophotonic |
|---|---|
| E | $E_{total} > \dfrac{\hbar\omega}{\eta_L \eta_D \eta_M \eta_C} \cdot \dfrac{V_r C_d}{e} \cdot 10^{\frac{\alpha L}{10}} + \dfrac{V_m \Theta \Delta T}{\left(\dfrac{dn}{d\rho}\right)\left(\dfrac{dT}{dn}\right)} +$ $\dfrac{2}{B} P_{tune} \times \Delta\lambda + E_{SD}\dfrac{B}{2F_{clock}}$ |
| β | $\dfrac{NB}{0.12 \log_e\left(\dfrac{56.6L}{\pi}\right)}$ |
| f | $< \dfrac{v_{sat}}{nw}$ |
| D | $< \dfrac{v_{sat}}{nk}\left(\dfrac{\lambda}{2N}\right)^{-3}$ |

**Table 2: Device Requirements for sub 100 fJ/bit CMOS Nanophotonic Interconnects***

| Feature | Target | E.g. |
|---|---|---|
| **Component Speed** | > 40 Gbit/s | 10-50 Gbit/s [20-35] |
| **WDM channels** (Number of channels/waveguide) | >8 | >4 [29, 30, 34] |
| **Modulator** (Switching Energy/bit) | <10 fJ/bit | <10 fJ/bit [23, 38] |
| **Detector** (Effective Capacitance & Quantum Efficiency) | 1 fF, > -1 dB @ 40 Gb/s | 2fF [31-35, 55] |
| **Operating Voltages, Current** (Modulator Drive and Detector Out) | ~ 600 mV (1.2 V differential), < 1 mA | 150 mV [38] |
| **Waveguide Losses** (High Confinement) | < 1 dB/cm | 6dB/cm [e.g. 56] |
| **Coupling Loss** (Single Mode Fiber to waveguide) | < 1dB | [e.g. 57] |
| **Laser Quantum Efficiency** | > -6 dB | -9 dB [e.g. 58] |
| **Serialization-Deserialization** | < 10 fJ/bit | see Appendix C |
| **Tuning Power** (@ 1nm/C change for low modal volume devices) | 100 µW/nm | 225 µW/nm [e.g. 42] |
| **Operating Range** | 20 K run-time | 50 K [e.g. 40] |

*We provide one possible set of device parameters. A large range of devices may meet the requirement with appropriate tradeoffs and appropriate scaling. The experimental devices typically demonstrate best performance only in one or few metric.*

*B. Compactness:* The dimensions of modulator, detector, switches and delays directly contribute to the areal density of interconnects and reduce the energy per bit. The target sizes of the modulators and detectors are less than 1 µm². Areal bandwidth density > 500 Tbit/mm².s, and footprint < 10 µm² are essential to meet the requirements of future interconnects.

*C. Multi-wavelength:* Multiple wavelength operation is essential for the linear interconnect density scaling. Wavelength Divison Multiplexing (WDM) is ideally suited for an on-chip optical interconnect due to complexity, foot print and optical insertion loss considerations.

*D. CMOS Compatibility:* The modulators, detectors, switches must operate with available voltage and current requirements of digital CMOS. Compatibility in drive currents and voltages must be ensured so that future technology nodes may allow for direct logic drive operation of the interconnect components

## VIII. CONCLUSION

We introduce an analytical framework for scaling nanophotonic interconnects to meet the energy and footprint requirements of CMOS global interconnects. We emphasize that the goal of this paper is to lay out a framework for a scaling path for optical devices and not provide a direct comparison with the several emerging promising technologies such as low swing voltage modulation. The adoption of any of the emerging technologies including photonic interconnects depends not only on the above figures of merits but on a combination of the HPC computing requirements, activity factors, cost, robustness to variations and noise margins. The following conclusions can be drawn for the photonic technology scaling requirements for CMOS global interconnects:

1. Scaling link bandwidth to 40 Gb/s and beyond can enable competitive energy/bit and areal bandwidth density. Improvement in link speed must be accompanied by improvement in SerDes operation.
2. Scaling the operational voltages of all electro-optics (< 0.6 V) to follow the CMOS voltage scaling is desirable.
3. Scaling the number of wavelengths per waveguide is essential to meet the linear bandwidth density of the global interconnects.
4. Fundamental limitations to the compactness of the optical devices may mandate 3D integration. If a viable 3D integration scheme does emerge, the photonic device layer may be unconstrained in area.
5. Improvement in thermal stability of the electro-optic detectors and modulators and passive elements beyond 10 µW/K is essential for stable operation of the links. The goal is to provide the performance with no change in the module level thermal management.
6. High conversion efficiency lasers (> 25%) & low insertion loss (< 8 dB) modulation, wave-guiding, and detection schemes are essential for low energy / bit operation.

With the appropriate scaling of device performance, photonic CMOS for on-chip interconnects may emerge as a technology for high performance computing applications in the CMOS/beyond-CMOS era.

### APPENDIX A: DERIVING OPTICAL LINK ENERGY

The energy per bit of the $E_{Source, Detector}$ can be derived as follows. At the detector end, the charge through the detector for 1 ON bit (and current) is given by

$$Q_{injected} = C_d V_r, \quad i_{detector} = C_d V_r B \quad (A.1)$$

The incident optical energy at the detector can be written as:

$$E = \frac{P_{detector}}{B} = \frac{\hbar\omega}{\eta_d} \frac{C_d V_r}{e} \quad (A.2)$$

Which gives the total electrical energy as:

$$E_{Source,Detect} > \frac{\hbar\omega}{\eta_L \eta_D \eta_M \eta_C} \cdot \frac{V_r C_d}{e} \cdot 10^{\frac{\alpha L}{10}} \quad (A.3)$$

### APPENDIX B: BIT ERROR CONSTRAINTS AT THE RECEIVER

For an N node interconnect network operating at frequency $f$, the tolerable error rate $P_{req}$ for operating with a failure rate of R over time T is [60]:

$$P_{req} < \frac{R}{NfT} \quad (B.1)$$

For 10,000 on chip global interconnects operating at 5 GHz with a failure rate of $10^{-6}$ over a lifetime of 10 years, the required error rate is $6.3 \times 10^{-29}$.

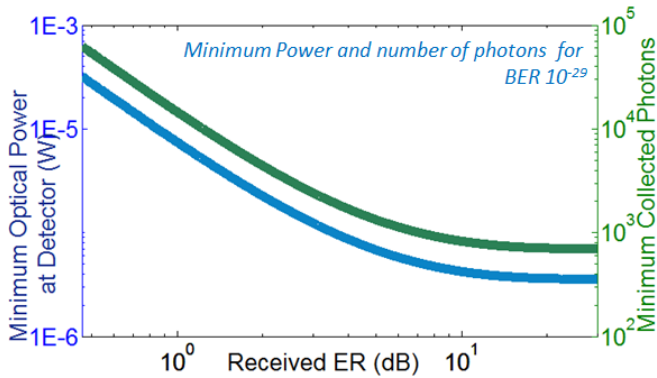

Figure B.1: Minimum optical power at a receiver less detector for BER $10^{-29}$

Following Beausolil et al [37], the mean number of photons required in an ON pulse for an error rate of P, for a modulation depth (1-M) for the off state of the light pulse is

$$n_{min} \approx \frac{-2\ln P}{\eta M^2}\left(2 - M + 2\sqrt{1 - M - \frac{M^2}{2\ln P}\frac{2kTC_d}{e^2}}\right) \quad (B.2)$$

Where $\eta$ is the total quantum efficiency of the detector. For a target error rate of $10^{-29}$, this corresponds to 823 collected photons per "ON" pulse at a detector capacitance 1 fF, modulation depth of M=0.9 (Extinction ratio=$-10\text{Log}_{10}(1-M)$= 10 dB).

The effect of modulation depth on the required optical power at the receiver (for a 40 Gbit/s signal) is shown in figure B.1. The minimum number of collected photons required at the given extinction ratio is also shown. Under the assumption of full charging of the detection capacitor (i.e. collected photons = $C_d V/e$=6240), we can see that the tolerable extinction ratio at the receiver is 1.4 dB. Hence, the degradation of the modulated optical signal due to insertion loss should not affect the BER for low insertion losses (< 8 dB). For the analysis of the paper we assumed that the modulators are maintained at optimal modulation depth using a tuning mechanism. We note that, the above received optical power is a lower limit for a receiver less detector. The degradation of SNR due to a TIA has to be accounted for in a receiver based system [63].

### APPENDIX C: EO MODULATOR ENERGY FOR ELECTRO-OPTIC POLYMER MODULATORS

A second common class of modulators compatible with CMOS is electro-optic polymer modulators [15]. The scaling with electro-optic properties for such modulators is as follows:

$$E_{EO} > C_m V_m^2 = \frac{C_m \Delta T^2}{\left(\frac{dT}{dn}\right)^2 \chi^2} \quad C.1$$

where $C_m$ is the modulator capacitance, $\Delta T$ is the modulation depth at the modulator and $\chi$ is the voltage electro-optic coefficient. The square law dependence with $\chi$ and $\Delta T$ are in contrast with carrier injection modulators.

### APPENDIX D: SCALING ESTIMATE FOR SERDES ENERGY

We arrived at an energy/bit/N (N=order of the SerDes multiplexing) scaling estimate assuming equal time performance at a given node. For equal time response, the ratio of the total channel width of the SerDes circuit is (for 32 nm CMOS vs. 11 nm CMOS):

$$r_w = \frac{\sum W_{11}}{\sum W_{32}} = \frac{C_{g11} V_{11} J_{32}}{C_{g32} V_{32} J_{11}} \approx 0.2696 \quad D.1$$

The ratio of the energy/bit/N can be estimated as:

$$r_E = \frac{C_{g11} V_{11}^2 \sum W_{11}}{C_{g32} V_{32}^2 \sum W_{32}} \approx 0.0797 \quad D.2$$

Equations D.1 and D.2 use the following values from ITRS 2011, PIDS2 HP CMOS table [1].

| Symbol | Parameter | 32 nm | 11nm (MG) |
|---|---|---|---|
| $C_g$ (fF/μm) | Ideal Gate Capacitance | $C_{g32}$ = 0.658 | $C_{g11}$= 0.338 |
| V (V) | HP Power supply | $V_{32}$ = 0.87 | $V_{11}$ = 0.66 |
| J (μA/μm) | NMOS drive current | $J_{32}$ = 1367 | $J_{11}$ = 1976 |

The estimated SerDes power at 32 nm under a global on chip synchronous clock without clock recovery is 27 fJ/bit/N [53]. Using the projected scaling ratio of 0.0797, at 11 nm node the estimated energy/bit/order is 2.16 fJ/bit/N (non-ideal gate capacitance as predicted by ITRS increases this projected value to 3.35 fJ/bit/N). To study the effect of the SerDes we have included a wide range of energy estimates of 100 fJ/bit/N to 10 fJ/bit/N in this paper.

APPENDIX E: EFFECT OF INSERTION LOSSES

The energy/bit of the optical interconnect is affected by the insertion losses due to the passive and active optical components. The insertion losses may arise from non-resonant modulator loss, mux, de-mux filters, waveguide crossing losses. The change in energy/bit due to total insertion losses is shown in figure E.1. Insertion losses can also play a major role if the degradation in extinction ratio at the detector reduces the received extinction ratio at the detector below the threshold for high bit error rate. For example, in section D, if the extinction ratio (of the received bits) reduces below 1.4 dB due to insertion loss, the interconnect will be BER limited.(for a modulator ER of 10 dB this places a 8.6 dB limit on IL)

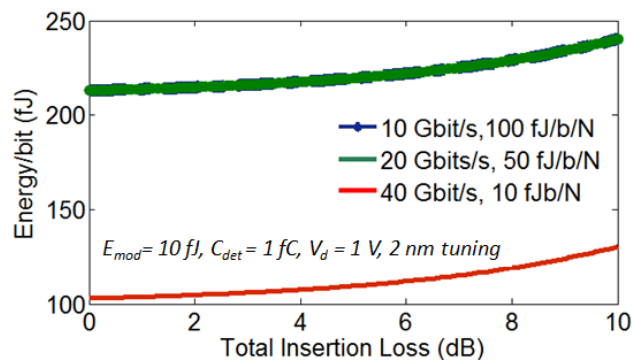

Figure E.1: Effect of insertion loss on the energy/bit

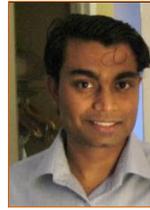

**Sasikanth Manipatruni (M'07)** is a research scientist in the Exploratory Integrated Devices and Circuits group in Intel Components Research. He is working on emerging novel devices to identify beyond CMOS logic technologies. He obtained his Ph.D. from Cornell University in silicon photonics. During his PhD, he was co-inventor of several silicon photonic devices including the first 18 Gb/s micro-ring modulator, first GHz poly-silicon modulator, EO switches & silicon nanophotonic links. He graduated from Indian Institute of Technology (IIT) Delhi at the top of his class in EE. He was a KVPY national science fellow of the Indian Institute of Sciences (IISc) (1999-2001); worked at Swiss Federal Institute of Technology (ETH), Zurich (2004) and Inter University center for astronomy and astrophysics (IUCAA) in 2001. He has 20 patent applications in nano-photonics, MRI, spintronics & 50 peer reviewed journal and conference papers. He serves as a peer reviewer for OSA, IEEE and Nature Photonics.

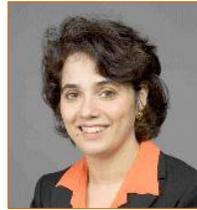

**Michal Lipson (SM'07)** received the B.S., M.S., and Ph.D. degrees in physics in the Technion—Israel Institute of Technology, Haifa, Israel, in 1998. In December 1998, she joined the Department of Material Science and Engineering, Massachusetts Institute of Technology (MIT) as a Postdoctoral Associate. In 2001, she joined the School of Electrical and Computer Engineering, Cornell University, where she is currently an Associate Professor. Her research at Cornell involves novel on-chip nanophotonic devices. She is the inventor of 12 patents regarding novel micron-size photonic structures for light manipulation. She is the author or coauthor of more than 100 papers in the major research journals in physics and optics.Dr. Lipson is a McArthur Fellow, Fellow of the Optical Society of America. She was the recipient of the National Science Foundation (NSF) CAREER Award, IBM Faculty Award and Blavatnik award, NY state academy of science .

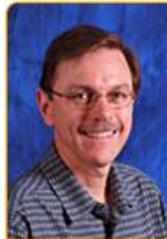

**Ian. A. Young** (M'78–SM'96–F'99) is a Senior Fellow and Director of Exploratory Integrated Circuits group in Intel Components Research. He is responsible for defining future circuit directions with emerging novel devices and identifying leading options for devices and interconnects to manufacture solid-state integrated circuits in the post-CMOS era. Dr. Young joined Intel in 1983. Starting with the development of circuits for a 1 Mb DRAM, and the world's first 1 µm 64 K SRAM, he then led the design of three generations of SRAM products and manufacturing test vehicles, and developed the original Phase Locked Loop (PLL) based clocking circuit in a microprocessor while working on the 50 MHz Intel 486™ processor design. He subsequently developed the core PLL clocking circuit building blocks used in each generation of Intel microprocessors through the 0.13 µm 3.2 GHz Pentium 4.

Born in Melbourne, Australia, he received his bachelor's and master's degrees in electrical engineering from the University of Melbourne, Australia. He received his Ph.D. in electrical engineering from the University of California, Berkeley in 1978. Prior to Intel, Dr. Young worked on analog/digital integrated circuits for Telecom products at Mostek Corporation (United Technologies), as well as an independent design consultant. Dr. Young was a member of the Symposium on VLSI Circuits Technical Program Committee from 1991 to 1996, serving as the Program Committee Chairman in 1995/1996, and the Symposium Chairman in 1997/1998. He was a member of the ISSCC Technical Program Committee from 1992 to 2005, serving as the Digital Subcommittee Chair from 1997 to 2003, the Technical Program Committee Vice-chair in 2004 and Chair in 2005. He was Guest Editor for the April 1997, April 1996 and December 1994 issues of the JSSC. He has served as an elected member of the SSCS Adcom from 2006 to 2009. Young received the International Solid-State Circuits Conference's 2009 Jack Raper Award for Outstanding Technology-Directions. Dr. Young is a Fellow of the IEEE. He holds 45 patents in integrated circuits and has authored or co-authored over 40 technical papers.